\documentclass[pra,twocolumn,superscriptaddress,showpacs]{revtex4}
\usepackage{amsmath,amssymb,bm,epic,eepic}
\usepackage{dsfont}

\newcommand{\beq}{\begin{equation}}
\newcommand{\enq}{\end{equation}}
\newcommand{\Tr}{\rm Tr}
\newcommand{\up}{\uparrow}
\newcommand{\down}{\downarrow}
\newcommand{\ket}[1]{|  #1  \rangle }

\newcommand{\bra}[1]{ \langle  #1  |}

\newcommand{\av}[1]{\langle #1 \rangle}
\newcommand{\nn}{\nonumber}
\newcommand{\forget}[1]{}
\newcommand{\half}{\frac{1}{2}}
\renewcommand{\vec}{\bm}

\renewcommand{\(}{\left(}
\renewcommand{\)}{\right)}
\newcommand{\goesto}{\longrightarrow}
\def\openone{\leavevmode\hbox{\normalsize1\kern-3.8pt\large1}}
\newcommand{\1}{\mathds{1}}
\newcommand{\BEQ}{\begin{eqnarray}}
\newcommand{\ENQ}{\end{eqnarray}}
\newcommand{\mbb}{\mathbb}

\newcommand{\C}{\ensuremath{\mbb{C}}}

\newcommand{\equivto}{\Longleftrightarrow}

\begin{document}
\title{Strengthened Bell inequalities for orthogonal spin directions}
\author{Jos Uffink} \email{uffink@phys.uu.nl}
\author{Michael Seevinck} \email{seevinck@phys.uu.nl}
\affiliation{%
Institute for History and Foundations of Science,\\
 Utrecht University
 PO Box 80.000, 3508 TA Utrecht, the Netherlands}%

\begin{abstract}\noindent
We strengthen the bound on the correlations of
 two spin-${1}/{2}$ particles (qubits) in separable (non-entangled) states for locally
orthogonal spin directions  by much tighter bounds than the well-known Bell
inequality.  This provides a sharper criterion for the experimental distinction
between entangled and separable states, and even one which is a necessary and
sufficient condition for separability.
 However, these improved bounds do not apply to
local hidden-variable theories, and hence they provide a criterion to test the
correlations allowed by local hidden-variable theories against those allowed by
separable quantum states.  Furthermore, these bounds are stronger than some
recent alternative experimentally accessible entanglement criteria. We also
address the issue of finding a finite subset of these inequalities that would
already form a necessary and sufficient condition for non-entanglement. For
mixed state we have not been able to resolve this, but for pure states a set of
six inequalities using only three sets of orthogonal observables is shown to be
already necessary and sufficient for separability.
\end{abstract}\pacs{03.65.Ud}

\maketitle

\section{Introduction}
\noindent
The current interest in the study of entangled quantum states derives from two
sources: their role in the foundations of quantum mechanics \cite{entanglement} and
their applicability in practical problems of information processing (such as quantum communication and computation) \cite{nielsen}.

Bell inequalities likewise serve a dual purpose. Originally, they
were designed in order to answer a foundational question: to test
the predictions of quantum mechanics against those of a local
hidden-variable (LHV) theory \cite{BELL64}. However, these
inequalities also provide a test to distinguish entangled from
separable quantum states \cite{GISIN91,H3}. Indeed, experimenters
routinely use violations of a Bell inequality to check whether
they have succeeded in producing entangled states. This problem of
entanglement detection is crucial in all experimental applications
of quantum information processing.

It is the goal of this letter to report that in the case of the standard Bell
inequality experiment, i.e., for two distant spin-${1}/{2}$ particles,
significantly stronger inequalities hold for separable states in the case of
locally orthogonal observables. These inequalities provide sharper tools for
entanglement detection, and are readily applicable to recent experiments such as \cite{volz}.  In fact,
if they hold for all sets of locally orthogonal observables they are necessary and sufficient for
separability, so the violation of these separability inequalities is not only a sufficient but also
a necessary condition for entanglement.
 They furthermore advance upon the necessary and sufficient separability inequalities of
Yu \emph{et al.} \cite{hefei}, since, in contrast to these, the inequalities
presented here do not need to assume that the orientations of the
measurement basis for each qubit are the same, so no shared reference frame is necessary.

We show the strength and efficiency of the separability criteria by showing that they are stronger than other sufficient and experimentally accessible criteria for
two-qubit entanglement while using the same measurement settings. These are (i) the so-called fidelity
criterion \cite{sackett}, and (ii) recent linear and non-linear
entanglement witnesses \cite{yu,nonlinear,zhang}.
However, in order to implement all of the above criteria successfully, the observables have to be chosen
in a specific way which depends on the state to be detected. So in general one needs some prior knowledge
about this state. In order to circumvent this experimental drawback we discuss the problem of whether
a finite subset of the separability inequalities could already provide a necessary and sufficient condition for
separability. For mixed state we have not been able to resolve this, but for pure states a set of six inequalities using only
three sets of orthogonal observables is shown to be already necessary and sufficient for separability.

The inequalities, however, are not applicable to the original purpose of
testing LHV theories. This shows that the purpose of testing
 entanglement within quantum theory, and the purpose of testing quantum
mechanics against LHV theories  are not equivalent, a point already
demonstrated by Werner \cite{WERNER89}. Our analysis follows up on Werner's
observation by showing that the correlations achievable by all separable
quantum states in a standard Bell experiment are tied to a bound strictly less
than those achievable for LHV models. In other words, quantum theory needs
entangled states even to produce the latter type of correlations. As an
illustration, we exhibit a class of entangled states, including the Werner
states, whose correlations in the standard Bell experiment possess a
reconstruction in terms of a local hidden-variable model.

This paper is organized as follows.  In section 2, we rehearse the Bell
inequalities  for separable states in the standard setting and derive a
stronger bound for orthogonal observables. In section 3, we compare this result
with that of LHV theories and argue that the stronger bound
does not hold in that case. In section 4, we return to quantum theory and
derive an even stronger bound than in section 2 which provides a necessary and sufficient criterion for
separability of all quantum states, pure or mixed. Furthermore, it is shown
that the orientation of the measurement basis is not relevant for the criterion
to be valid. Section 5 compares the strength of these inequalities with some
other criteria for separability, not based on the Bell inequalities. Also, it is investigated whether
a finite subset of the inequalities of section 4 could already provide a necessary and  sufficient separability condition.
Section 6 summarizes our conclusions.

\section{Bell inequalities as a test for entanglement}
\noindent Consider a system composed of a pair of spin-${1}/{2}$
particles (qubits) on the Hilbert space $\mathcal{H}=\mathbb{C}^2 \otimes \mathbb{C}^2$  in the familiar setting of a standard Bell
experiment consisting of two distant sites, each receiving one of
the two particles, and where, at each site, a choice of measuring
either of two spin observables is made.
 Let $A,
A'$ denote the  two spin observables on the first particle, and $B, B'$ on the
second. We write $AB$ etc., as shorthand for $A\otimes B$ and $\av{AB}_\rho :=
\mbox{Tr}[\rho A\otimes B]$ or $\av{A B}_\Psi = \bra{\Psi}A \otimes B
\ket{\Psi}$ for the expectations of $AB$ in the mixed state $\rho$ or pure
state $\ket{\Psi}$.

It is well known that for all such observables and all separable
states, i.e., states of the form $\rho= \rho_1\otimes\rho_2$ or
convex mixtures of such states, the Bell
 inequality in the form derived by Clauser, Horne, Shimony and Holt (CHSH) \cite{chsh}
holds:
 \begin{equation} |\langle AB + AB' + A'B -
A'B'\rangle_\rho|  \leq 2.   \label{1} \end{equation} The maximal
violation of (\ref{1})
 follows from an inequality of Cirel'son~\cite{CIRELSON}
 (cf.\
 Landau~\cite{LANDAU}) that holds for all quantum states:
 \begin{equation} |\langle AB + AB' +
A'B - A'B'\rangle_\rho | \leq 2 \sqrt{2}. \label{2}\end{equation}
Equality in (2) ---and thus the maximal violation of inequality
(1) allowed in quantum mechanics--- is attained by e.g. the pure
entangled  states
$\ket{\phi^{\pm}}=\frac{1}{\sqrt{2}}(\ket{\up\up} \pm
\ket{\down\down})$ and
$\ket{\psi^{\pm}}=\frac{1}{\sqrt{2}}(\ket{\up\down} \pm
\ket{\down\up})$.

In Ref.\ \cite{UFFINK02} it was furthermore shown that for all such observables and for all states $\rho$
\begin{equation}
\av{AB' + A'B }_\rho^2 + \av{A B - A' B'}_\rho^2 \leq 4,
\label{3}\end{equation}
which strengthens the Cirel'son inequality (2). This quadratic inequality (3) is
likewise saturated for maximally entangled states like $\ket{\psi^\pm}$ and
$\ket{\phi^\pm}$.
 Unfortunately, no smaller
bound  on the left-hand side of (\ref{3}) exists for separable
states, as long as the choice of observables is kept  general. (To
verify this, take $\ket{\Psi} = \ket{\up\up}$ and $A =A'=B=B' =
\sigma_z$) Thus, the quadratic inequality (\ref{3}) does not
distinguish entangled and separable states.  We now show that a
much more stringent bound can be found  on the left-hand side of
(3) for separable states when a suitable choice of observables is
made, exploiting an idea made in a different context by
\cite{TOTHSEEV}.

For the case of the singlet state $\ket{\psi^-}$, it has long been
known that an optimal choice of the spin observables for the
purpose of finding violations of the Bell inequality requires that
$A,A'$ and $B,B'$ are pairwise orthogonal, and many experiments
have chosen this setting. And for general states, it is only in
such locally orthogonal configurations that one can hope to attain
equality in inequality (\ref{2}) \cite{popescu,wernerwolf,seevuff2007}. It is  not true,
however, that for any given state $\rho$ the maximum of the left
hand side of the Bell-CHSH inequality always requires orthogonal
settings \cite{GISIN91,H3,POPESCUROHRLICH}.

However this may be, we will from now on assume local orthogonality,
 i.e., $A\perp A'$  and $B\perp
B'$ (for the case of two qubits this amount to the local observables
anticommuting with eachother: $\{A,A'\}=0=\{B,B'\}$). Furthermore, assume for
the moment that the two-particle state is pure and separable. We may thus write
$\rho = \ket{\Psi}\bra{\Psi}$, where $\ket{\Psi} = \ket{\psi} \ket{\phi}$, to
obtain:
\begin{align} \av{AB' + A'B }_\Psi^2 +& \av{A B - A' B'}_\Psi^2 =
 \(\av{A}_\psi\av{B'}_\phi + \av{A'}_\psi\av{B}_\phi \)^2 \nn\\
 & +  \(
\av{A}_\psi\av{ B}_\phi - \av{ A'}_\psi \av{ B'}_\phi\)^2 \nn \\
=&\( \av{A}_\psi^2 + \av{A'}_\psi^2 \) \(\av{B}_\phi^2 + \av{B'}_\phi^2\).  \end{align}
Now, for any pure spin-$\half$ state $\ket{\psi}$ on $\mathcal{H}=\mathbb{C}^2$, and any orthogonal triple of
spin components $A, A'$ and $A''$, one has $\av{A}_\psi^2 +
\av{A'}_\psi^2 + \av{A''}_\psi^2 =1$, and similarly $\av{B}_\phi^2
+ \av{B'}_\phi^2 + \av{B''}_\phi^2=1$. Therefore, we can write (4)
as: \beq \av{AB' + A'B }_\Psi^2 + \av{A B - A' B'}_\Psi^2  =
\left(1 - \av{A''}_\psi^2\right)\left(1 - \av{B''}_\phi^2\right) .
\label{newa} \enq \forget{As a result, the left hand side of
(\ref{3}) is bounded by $1$ for any pure product state.} This
result for pure separable states can be extended to any  mixed
separable state by noting that the density operator of any such
state is a convex combination of the density operators for pure
product-states, i.e. $\rho = \sum_i p_i \ket{\Psi_i}\bra{\Psi_i}$,
with $\ket{\Psi_i} =\ket{\psi_i} \ket{\phi_i}$,  $p_i\geq 0$ and
$\sum_i p_i=1$.  We may thus write for such states:
\begin{align}   & \av{AB' + A'B }_\rho^2 + \av{A B - A' B'}_\rho^2 \nn\\
& \leq\(\sum_{i} p_i  \sqrt{\av{AB' + A'B }^2_i  +  \av{A B - A'B'}^2_i } \)^2
\nn\\  & =   \(\sum_i p_i \sqrt{ \( 1 - \av{A''}^2_i\) \( 1 -
\av{B''}^2_i\)}\)^2  \nn\\ & \leq \( 1 - \av{A''}^2_ \rho\) \( 1 -
\av{B''}^2_\rho\).
\end{align}
Here, $\av{\cdot}_i$ denotes  an expectation value with
respect to $\ket{\Psi_i}$ (e.g.,
$\av{A''}_i:=\bra{\Psi_i} A''\otimes\1\ket{\Psi_i}$) and
$\av{A''}_\rho:=\av{A''\otimes\1}_\rho$. The first inequality follows because
  $\sqrt{\av{AB' + A'B }^2_\rho  +  \av{A B - A'B'}^2_\rho }$ is a
convex function of $\rho$  and the second
because $\sqrt{ \( 1 - \av{A''}^2_ \rho\) \( 1 - \av{B''}^2_\rho\)}$ is concave
in $\rho$. (To verify this, it is helpful to use the general lemma that for all
positive concave functions $f$ and $g$, the function $\sqrt{fg}$ is concave.)

 Thus, we obtain for all separable states and
locally orthogonal triples $A\perp A'\perp A''$, $B\perp B'\perp
B''$: \begin{align} \av{AB' + A'B }_\rho^2 + \av{A B - A' B'}_\rho^2 \leq
\left(1 - \av{A''}_\rho^2\right)\left(1 - \av{B''}_\rho^2\right)
. \label{new} \end{align}
Clearly, the right-hand side of this inequality is bounded by 1. However, as
noted before, entangled states can saturate inequality (\ref{3}) ---even for
orthogonal observables--- and attain the value of 4  for the left-hand side and
thus clearly violate the bound (\ref{new}). In contrast to (3), inequality
(\ref{new}) thus does provide a criterion for testing entanglement.    The
strength of this bound for entanglement detection in comparison with the
traditional Bell-CHSH inequality (\ref{1}) may be illustrated by considering
the region of values it allows  in the $(\av{X}_\rho,\av{Y}_\rho)$-plane, where
$\av{X}_\rho = \av{AB - A'B' }_\rho$ and $\av{Y}_\rho = \av{AB' + A'B }_\rho$,
cf.\ Fig.~1. Note that even in the weakest case, (i.e., if $\av{A''}_\rho=
\av{B''}_\rho=0$), it wipes out just over 60\% of the area allowed by
inequality (\ref{1}). This quadratic inequality even implies
 a strengthening of the Bell-CHSH inequality (\ref{1}) by a factor
$\sqrt{2}$:
\begin{equation} |\langle AB + AB' + A'B -
A'B'\rangle_\rho|  \leq \sqrt{2},   \label{sqrt2}
\end{equation} recently
obtained by Roy \cite{roy}. In fact, even if one chooses only one pair (say
$B,B'$) orthogonal, and let $A,A'$ be arbitrary, one would obtain an upper
bound of 2 in (\ref{new}), and still improve upon the Bell-CHSH inequality.
Another pleasant feature of inequality (\ref{new}) is that for pure states it's
violation is a necessary and sufficient condition for entanglement (see
Appendix A). Also, for future purposes we note that the expression in left-hand
side is invariant under rotations of $ A, A'$ around the axis $A''$ and
rotations of $B,B'$ around $B''$.\vskip0.5cm

 \setlength{\unitlength}{0.08 mm}
 \begin{figure}[h]\begin{center}
\begin{picture}(200,510)(-100,-180)
\forget{ \put(-50,-50){\line(0,1){100}}
 \put(-50,-50){\line(1,0){100}}
  \put(50,50){\line(0,-1){100}}
 \put(50,50){\line(-1,0){100}}}
 \put(0,-200){\line(1,1){200}}
 \put(0,-200){\line(-1,1){200}}
  \put(0,200){\line(1,-1){200}}
 \put(0,200){\line(-1,-1){200}}
 \put(0,0){\circle{200}}
 \put(0,0){\circle{400}}
 \put(310,0){\makebox(0,0){$\av{X}_\rho$}}
 \put(10,293){\makebox(0,0){$\av{Y}_\rho$}}
 \put(0,230){\makebox(0,0){$2$}}
  \put(230,0){\makebox(0,0){$2$}}
   \put(0,-230){\makebox(0,0){$-2$}}
    \put(-239,0){\makebox(0,0){$-2$}}
\end{picture}\end{center}\vspace{\baselineskip}
\caption{\it \small Comparing the regions in the
$(\av{X},\av{Y})$-plane allowed (i) by inequality (\ref{3}) which
holds for all quantum states (the inside of the large circle);
(ii) by
 the \mbox{Bell}-CHSH inequality (\ref{1}) (the inside of the square); and (iii) by the strengthened inequality
(\ref{new}) which holds for all separable quantum states (inside
the inner circle with radius $1$).}
\vspace{\baselineskip}
\end{figure}
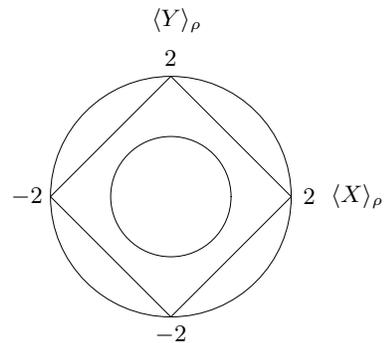

The inequalities (\ref{new}) present a necessary criterion for a
quantum state to be separable ---and its violation thus a
sufficient criterion for entanglement---, but in contrast to pure
states,  they are  clearly not sufficient for separability  of mixed
states. In section  \ref{necsuf} we shall present an even stronger
set of inequalities that is necessary and sufficient for mixed states as
well,  but we will first discuss the results obtained so far  in the light of
LHV theories.

\section{Comparison to local hidden-variable theories}
\noindent
It is interesting to ask whether one can obtain a similar stronger
inequality as  (\ref{new}) in the context of local hidden-variable
theories. It is well known that inequality (\ref{1})  holds also
for any such theory in which dichotomous outcomes $a,b \in \{+, -\}$
are subjected to a probability distribution \beq p(a,b) =
\int_\Lambda \!d\lambda\, \rho(\lambda) P_{\vec{a}}(a|\lambda)
P_{\vec{b}} (b|\lambda).\enq Here, $\lambda\in \Lambda$ denotes the
``hidden variable'', $\rho(\lambda)$ denotes a probability density
over $\Lambda$, $\vec{a}$ and $\vec{b}$ denote the `parameter settings', i.e., the directions of
the spin components measured, and $P_{\vec{a}}(a|\lambda)$, $P_{\vec{b}}(b|\lambda)$
are the probabilities (given $\lambda$) to obtain outcomes $a$ and $b$ when measuring the settings
$\vec{a}$ and $\vec{b}$ respectively. The locality condition is expressed by the
factorization condition $P_{\vec{a}, \vec{b} } (a,b|\lambda)
=P_{\vec{a}}(a|\lambda) P_{\vec{b}} (b|\lambda)$.

The assumption to be added to such an LHV theory in order to obtain
the strengthened inequality (\ref{new}) is the requirement that for any
orthogonal choice of $A, A'$ and $A''$ and for every given $\lambda$
we have \beq
\av{A}_{\lambda}^2 + \av{A'}_{\lambda}^2  + \av{A''}_\lambda^2 = 1,
\label{reqa}\enq or at least \beq \av{A}_{\lambda}^2 +
\av{A'}_{\lambda}^2 \leq 1 \label{req},\enq
where $\av{A}_{\lambda}
= \sum_{a=\pm1}  a\, P_{\vec{a}}\,(a|\lambda)$, \forget{$\av{A'}_{\rm hv}
= \int d \lambda \rho(\lambda) p_\vec{a'}(a'= +|\lambda -
p_\vec{a'}(a'= -|\lambda$; and similar for $\av{B}_{\rm
hv},\av{B'}_{\rm hv}$.} etc.

But a requirement like (\ref{reqa}) or (\ref{req}) is by no means
obvious for a local hidden-variable theory. Indeed, as has often
been pointed out, such a theory may employ a mathematical
framework which is completely different from quantum theory. There
is no \emph{a priori} reason why the orthogonality of spin
directions should have any particular significance in the
hidden-variable theory, and  why such a theory should confirm to
quantum mechanics in reproducing (\ref{req})  if one
conditionalizes on a given hidden-variable state. (One is reminded
here of Bell's critique \cite{BELL} on von Neumann's `no-go'
theorem.) Indeed, (\ref{req}) is violated by Bell's own example of
an LHV model \cite{BELL64} and in fact it must fail in every
deterministic LHV theory
(where all probabilities $P_{\vec{a}}(a|\lambda)$, $P_{\vec{b}}(b|\lambda)$ are either $0$ or $1$), since for those theories
$\av{A}_\lambda^2 =\av{A'}^2_\lambda = \av{A''}^2_\lambda =1$.
 Thus, the additional requirement (\ref{req})
would appear entirely  unmotivated  within an LHV theory.

It thus appears that testing for entanglement within quantum
theory and testing quantum mechanics against the class of all LHV
theories are not equivalent issues. Of course, this conclusion is
not new. Werner \cite{WERNER89} has already constructed an
explicit LHV model  for a specific entangled state. Consider the
so-called Werner states: $ \rho_W = \frac{1-p}{4} \1   + p
\ket{\psi^-}\bra{\psi^-}$,   $p\in[0,1]$. Werner showed \cite{WERNER89} that
these states are entangled if \mbox{$p> {1}/{3}$}, but
nevertheless possess an LHV model for \mbox{$p= {1}/{2}$}.  The
above inequality (\ref{new}) suggests that the phenomenon
exhibited by this Werner state is much more ubiquitous, i.e., that
many more entangled states have an LHV model. We will show that
this is indeed the case.

 It is not easy to find the general set of quantum states that
 possess an LHV model \cite{WERNER89, acin}. Certainly, the question cannot be decided by considering orthogonal observables only.
However, as shown in Appendix B, it is possible to determine the class of states for which
 \begin{align}  \max_{A\perp A',
B\perp B'} \av{AB' + A'B }_\rho^2 + \av{A B - A' B'}_\rho^2 >1
\end{align} holds (they are thus entangled), and which in addition
satisfy the Bell-CHSH inequalities of \mbox{Eq. (\ref{1})} for
\emph{all} choices of observables, i.e., not restricted to
orthogonal directions.

Since the latter are known \cite{wernerwolf2,zukowskibrukner,fine,pitowsky}
to form a necessary and sufficient set of conditions for the existence
of an LHV model for all standard Bell experiments on spin-${1}/{2}$ particles,
 we conclude that all
correlations obtained from such entangled states can be
 reconstructed by an LHV model \cite{footnote}.
It  follows from Appendix B that  this class of states includes
the Werner states for the region ${1}/{2} < p \leq {1}/{\sqrt2}$,
which complements results obtained by Ref. \cite{H3} in which the
non-existence of an LHV model is demonstrated for ${1}/{\sqrt2}< p
\leq1$.

\section{A necessary and sufficient condition for separability \label{necsuf}}\noindent The inequalities
(\ref{new}) can be strengthened even further. To see this it is
useful to introduce, for some given pair of locally orthogonal
triples $(A, A', A'')$ and $(B,B',B'')$, eight new two qubit operators on $\mathcal{H}=\mathbb{C}^2\otimes\mathbb{C}^2$:
\BEQ
I := \half (\1 + A'' B'') &~~~~~~~&
 \tilde{I} := \half (\1 - A'' B'' )
 \nn\\
X:=   \half (A B - A' B' )
&~~~~~~~& \tilde{X}:=   \half (A B  + A' B' )\nn\\
Y:=   \half (A'B  +   AB'  )
&~~~~~~~&\tilde{Y}:=   \half (A' B  -  AB' ) \nn\\
 Z :=  \half (A'' + B'' ) &~~~~~~~& \tilde{Z} :=  \half (  A'' -  B'' )\, ,
 \nn\\
   \ENQ
where   $ \half (A'' + B'' )$ is shorthand for $\half(A''\otimes\1 +\1\otimes B'')$, etc.
Note that $X^2 = Y^2 = Z^2 = I^2 =I$ and similar for their tilde versions,
  and that all eight operators mutually anti-commute.
Furthermore, if the orientations of the two triples is the same (e.g., $[A, A']=2iA''$ and $[B, B']=2iB''$), they form
   two representations of the generalized
Pauli-group, i.e. they have the same commutation relations as the
Pauli matrices on $\C^2$, i.e.: $ [X,Y]=2iZ$, etc., and $\av{X}_\rho^2+\av{Y}_\rho^2+\av{Z}_\rho^2\leq\av{I}_\rho^2$ with equality only for pure states (analogous for the tilde version).  Note that
these two sets transform in each other by replacing $B' \goesto -
B'$ and $B'' \goesto - B''$.

Now we can repeat the argument of section 2. Let us first
temporarily assume the state to be pure and separable, $\ket{\Psi}
=\ket{\psi}\ket{\phi}$.  We then obtain:
\begin{align}
\av{X}_\Psi^2 +
\av{Y}_\Psi^2 &= \frac{1}{4} \left( \av{AB -A'B'}_\Psi^2 +
\av{A'B + A B'}_\Psi^2\right) \nn\\
&=\frac{1}{4} \left( \av{A}_\psi^2 + \av{A'}_\psi^2 \right) \left(\av{B}_\phi^2 + \av{B'}_\phi^2\right) \nn\\
&= \av{\tilde X}_\Psi^2 + \av{\tilde Y}_\Psi^2 \label{XY}
\end{align}
and similarly: \begin{align}
\av{I}_\Psi^2 - \av{Z}_\Psi^2& = \frac{1}{4} \left( \av{1 + A'' B''}_\Psi^2 - \av{A'' + B''}_\Psi^2 \right)\nn\\
&= \frac{1}{4} \left( 1 -\av{A''}_\psi^2 \right) \left( 1 - \av{B''}_\phi^2 \right)\nn\\
&= \av{\tilde I}_\Psi^2  -\av{\tilde Z}_\Psi^2  \label{IZ}.\end{align}
In view of (\ref{newa}) we conclude that for all pure separable
states all expressions in the equations (\ref{XY}) and (\ref{IZ})
are equal to each other.   Of course, this conclusion does not
hold for mixed separable states. However,  $\sqrt{\av{ X}_\rho^2 +
\av{ Y}_\rho^2}$ and   $ \sqrt{\av{\tilde X}_\rho^2 + \av{\tilde
Y}_\rho^2}$ are convex functions of $\rho$ whereas
 the three expressions $\sqrt{\av{I}_\rho^2 - \av{Z}_\rho^2}$, $\frac{1}{4}\sqrt{ \left(
1 -\av{A''}_\rho^2 \right)
 \left( 1 - \av{B''}_\rho^2\right)}$ and $\sqrt{\av{\tilde I}_\rho^2 - \av{\tilde Z}_\rho^2}$
 are all concave in $\rho$.
 Therefore we can repeat
 a similar
 chain of reasoning as in (6) to  obtain  the following inequalities, which are valid
for all mixed separable states: \begin{align} \left.
\begin{array}{c}
\av{X}_\rho^2 + \av{Y}_\rho^2 \\
\av{\tilde{X}}_\rho^2 + \av{\tilde{Y}}_\rho^2 \end{array} \right\}
\leq
\left\{ \begin{array}{c} \av{\tilde I}_\rho^2 - \av{\tilde Z}_\rho^2 \\
\frac{1}{4} \left( 1 -\av{A''}_\rho^2 \right)
 \left( 1 - \av{B''}_\rho^2\right)\\
 \av{I}_\rho^2 - \av{Z}_\rho^2 \end{array}
 \right. .\label{mixsep} \end{align}

This result extends the previous inequality  (\ref{new}).
  The next obvious  question is then which of the three  right-hand
sides in (\ref{mixsep}) provides  the lowest  upper bound. It is
not difficult to show that the ordering of these three expressions
depends on the correlation coefficient $C_\rho = \av{A''B''}_\rho
- \av{A''}_\rho\av{B''}_\rho$. A straightforward calculation shows
that if $C_\rho \geq 0$, \begin{align} \av{I}^2_\rho - \av{Z}^2_\rho \leq \frac{1}{4}
\left( 1 -\av{A''}^2_\rho \right)
 \left( 1 - \av{B''}^2_\rho\right)  \leq \av{\tilde I}^2_\rho - \av{\tilde Z}^2_\rho \end{align}
 while the above inequalities are inverted  when $C_\rho \leq 0$.  Hence, depending on the sign of $C_\rho$, either $\av{I}^2_\rho
-\av{Z}^2_\rho$ or $\av{\tilde I}^2_\rho - \av{\tilde{Z }}^2_\rho$ yields the
sharper upper bound. In other words, for all separable quantum
states one has:
\begin{align} \left.
\begin{array}{c}
\av{X}_\rho^2 + \av{Y}_\rho^2 \\
\av{\tilde{X}}_\rho^2 + \av{\tilde{Y}}_\rho^2 \end{array} \right\}
\leq
\left\{ \begin{array}{c} \av{\tilde I}_\rho^2 - \av{\tilde Z}_\rho^2 \\
 \av{I}_\rho^2 - \av{Z}_\rho^2 \end{array}
 \right. .\label{mixsep2} \end{align}
This set of inequalities provides the announced strengthening of
(\ref{new}). This improvement pays off: in contrast to
(\ref{new}), the validity of the inequalities (\ref{mixsep2}) for all orthogonal triples $A,A',A''$ and $B,B',B''$ provides a necessary
and sufficient condition for separability for all states pure or
mixed. (See appendix C for a proof).

We note that a special case of the inequalities (\ref{mixsep2}), to wit
\beq \av{\tilde{X}}_\rho^2 + \av{\tilde{Y}}_\rho^2
\leq \av{I}_\rho^2 - \av{Z}_\rho^2  \label{hef} \enq was already found by Yu \emph{et
al.} \cite{hefei}, by a rather different argument. These authors stressed that
the orientation of the locally orthogonal observables play a crucial role in
this inequality: if one chooses both triples to have a \emph{different}
orientation (i.e., $A = i [A', A'']/2$ and $B = -i [B', B'']/2$ or $A =
-i [A', A'']/2$ and $B = i [B', B'']/2$) the inequality (\ref{hef}) holds
trivially for all quantum states $\rho$, whether entangled or not. It is only
when the orientation between those two triples is \emph{the same} that
inequality (\ref{hef}) can be violated by entangled  quantum states.

 The present result (\ref{mixsep2})
 complements their findings by showing that  the
relative orientation of the two triples is not a crucial factor in
entanglement detection. Instead, if the orientations are the same,
both of the following inequalities contained in (\ref{mixsep2}) \begin{align} \av{X}_\rho^2 +
\av{Y}_\rho^2 &\leq \av{\tilde{I}}_\rho^2 -
\av{\tilde{Z}}_\rho^2 \label{21} \\
\av{\tilde X}_\rho^2 + \av{\tilde Y}_\rho^2  &\leq
\av{{I}}_\rho^2 - \av{{Z}}_\rho^2 \label{22} \end{align} are useful
tests for entanglement, while the remaining two become trivial. If on the other hand, the orientations are opposite,
their role is taken over by \begin{align} \av{X}_\rho^2 + \av{Y}_\rho^2
&\leq \av{{I}}^2_\rho -
\av{{Z}}_\rho^2 \\
\av{\tilde X}_\rho^2 + \av{\tilde Y}_\rho^2 & \leq
\av{\tilde{I}}_\rho^2 - \av{\tilde{Z}}_\rho^2 \end{align} while
(\ref{21}) and (\ref{22}) hold trivially.

\section{Experimental aspects of the new inequalities} In this section we compare
the strength of the inequalities (\ref{mixsep2}) to some other experimentally feasible
 criteria to distinguish separable and entangled two qubit states that are not based on Bell-type inequalities.
 Also, we discuss the problem of whether a
finite set of triples for the inequalities (\ref{mixsep2}) could
be necessary and sufficient for separability.

A well-known alternative criterion for separability is the
fidelity criterion,  which says that for all separable states the
fidelity $F$ (i.e., the overlap with a Bell state
$\ket{\phi_\alpha^+} = \frac{1}{\sqrt{2}}(\ket{\up\up} +
e^{i\alpha}\ket{\down\down}$), $\alpha \in \mathbb{R}$) is bounded
as \begin{align}
 F (\rho) :&=\max_\alpha
 \bra{\phi_\alpha^+}\rho\ket{\phi_\alpha^{+}}\nn\\
&=\frac{1}{2}(\bra{\up\up} \rho \ket{\up\up} +\bra{\down\down}
\rho\ket{\down\down}) + |\rho_\nearrow |  \leq \frac{1}{2}. \label{fid}
\end{align}  Here, $\rho_\nearrow$ denotes the extreme anti-diagonal element of
$\rho$, i.e., $\rho_\nearrow=\bra{\up\up} \rho \ket{\down\down}$.  For a proof,
see \cite{sackett}. An equivalent formulation  of (\ref{fid}), using
$\mbox{Tr}\rho =1$ is
 \beq  2 | \rho_\nearrow|
\leq
   \bra{\up\down}\rho
\ket{\up\down} +  \bra{\down\up} \rho \ket{\down\up} \label{fid2}.\enq
   However, choosing the standard Pauli
matrices for   both triples, i.e. $(A, A', A'') =( B,B', B'') =
(\sigma_x, \sigma_y , \sigma_z)$ we obtain  from (\ref{mixsep2})
 \begin{align} &\av{X}^2 + \av{Y}^2 \leq  \av{\tilde I}_\rho^2 -
\av{\tilde Z}_\rho^2 ~~~~~ \equivto\nn \\ &4 |\rho_\nearrow|^2  \leq  \(
\bra{\up\down}\rho\ket{\up\down} +  \bra{\down\up} \rho \ket{\down\up} \)^2
-\nn\\& ~~~~~~~~~~~~~~~~~~~~~~~~~~~~~~~ \(  \bra{\up\down}\rho \ket{\up\down}-
\bra{\down\up} \rho \ket{\down\up} \)^2
\end{align} which implies  (\ref{fid2}). Clearly, then, (\ref{mixsep2}) is
stronger than the fidelity criterion, i.e., it will detect more entangled
states.

As another application,  consider
the following entanglement witnesses for so-called local orthogonal
observables (LOOs) $\{G_k^A\}_{k=1}^4$ and $\{G_k^B\}_{k=1}^4$: a linear one
presented by  \cite{yu}: \beq \av{\mathcal{W}}_\rho = 1- \sum_{k=1}^4  \av{G_k^A
\otimes G_k^B}_\rho, \enq and  a nonlinear witness from \cite{nonlinear}
given by \beq \mathcal{F}(\rho)=1-\sum_{k=1}^4\langle G_{k}^{A}\otimes
G_{k}^{B}\rangle_{\rho}-\frac{1}{2}\sum_{k=1}^4\langle
G_{k}^{A}\otimes \1-\1\otimes
G_{k}^{B}\rangle^{2}_\rho.\enq
Here, the set $\{ G_k^A\}_{k=1}^4$ is a set of four observables that form a basis
for all operators in the Hilbert space of a single qubit and which satisfy
orthogonality relations Tr$[G_kG_{k'}]=\delta_{kk'}$ ($k,k'=1,\dots 4$).
A typical complete set of LOOs is formed by any orthogonal triple of spin
directions  conjoined with the identity operator, i.e., in the notation of this paper,
 $\{ G_k^A\}_{k=1}^4 = \{ \1 ,  A, A', A''\}/\sqrt{2} $ and similarly    for $\{ G_k^B\}_{k=1}^4 $.

   These witnesses provide tests for entanglement in the sense that for all separable states
$   \av{\mathcal{W}}_\rho \geq  0 ,~ \mathcal{F}(\rho) \geq 0$ must hold
     and a violation of either of these inequalities is thus a sufficient condition for entanglement.
 An optimization procedure for the choice of LOOs in
these two witnesses is given in Ref. \cite{zhang}.

The strength of these two criteria has been studied for the noisy singlet state
introduced in Ref. \cite{nonlinear}: $\rho=p \ket{\psi^{-}} \bra{\psi^-}+(1-p)\rho_{sep}\,$,  where
$\ket{\psi^{-}}=$\mbox{$(\ket{\up\down}-\ket{\down\up})/\sqrt{2}$} is the singlet
state and the separable noise is $\rho_{sep}=$\mbox{$2/3 \ket{\up\up} \bra{\up\up} + 1/3\ket{\up\down}\bra{\up\down}$}. The
Peres-Horodecki criterion \cite{PPT} gives that this state $\rho$ is entangled for any
$p>0$. Under the complete set of LOOs
$\{-\sigma_{x},-\sigma_{y},-\sigma_{z},\1\}^{A}/\sqrt{2}$,
$\{\sigma_{x},\sigma_{y},\sigma_{z},\1\}^{B}/\sqrt{2}$, the linear
witness given above can detect the entanglement for all $p>0.4$
\cite{zhang}, and the nonlinear one detects the entanglement for
$p>0.25$ \cite{nonlinear}.
 Using the optimization procedure of Ref. \cite{zhang} the
optimal choice of LOOs for the linear witness can detect the
entanglement for all $p>0.292$, whereas the nonlinear witness
appears to be already optimal.

Using the same set of LOOs as above, the quadratic separability inequality (\ref{mixsep2})
detects the entanglement already for $p>0$ (i.e., as soon as the
state is entangled it can be detected), and it is thus stronger
than these two witnesses for this particular state.

As a final topic, we wish to point out that, in spite of the
strength of the inequalities (\ref{mixsep2}), they also have an
important drawback from an experimental point of view.  In order to
check their validity or violation one would have to measure for
\emph{all} locally orthogonal triples of observables, a task which
is obviously unfeasible since there are uncountably many of those.
Because of this one must generally gather some prior knowledge about
the state whose entanglement is to be detected, so that one can choose settings that allow for a violation.
It is therefore highly interesting to ask whether a finite collection of orthogonal triples could be found for which the
satisfaction of these inequalities would already provide a
necessary and sufficient condition for separability, since then such prior knowledge would no longer be necessary. Measuring
the finite collection of settings would then be always sufficient for entanglement detection, independent of the state to be detected.

We have performed an (unsystematic) survey of this problem. A
first natural attempt would be to consider the triples obtained
by permutations of the basis vectors. Thus, consider the set of three
 inequalities obtained by taking for both  triples $(A, A',
A'')$ and $(B, B' , B'')$  the  choices $\alpha=(\sigma_x, \sigma_y
, \sigma_z)$, $\beta=(\sigma_z, \sigma_y, \sigma_x)$ and $\gamma=(\sigma_z, \sigma_x, \sigma_y)$. (Other permutations do not
contribute independent inequalities.)

Under this choice,  (\ref{mixsep2}) leads to
 the six inequalities \begin{align} \av{X_k}^2 + \av{Y_k}^2
&\leq
\av{\tilde{I}_k}^2 - \av{\tilde{Z_k}}^2 , \\
\av{\tilde{X}_k}^2 + \av{\tilde{Y}_k}^2 &\leq
 \av{{I}_k}^2 -
\av{{Z_k}}^2 , \end{align} for $k= \alpha,\beta,\gamma$.

For a general pure state $\ket{\Psi} = a \ket{\up\up} + b \ket{\up\down} + c
\ket{\down \up} + d\ket{\down\down}$, the satisfaction of these inequalities
boils down to three equations:
\begin{align}   |ad|   &= |bc| , \\
|   (a+d)^2-  (b+c)^2 |   &=   |(a-d)^2   - (b-c)^2 |,\\
|(b+c)^2   +  (a-d)^2 | &= | (b-c)^2 +  (a+d)^2|. \end{align}
 However,  these equations are satisfied if $a =c =i$, $ -b= d =1$, i.e.
for an entangled pure state. This shows that the  choice
$\alpha$, $\beta$, $\gamma$ above does not produce a
 sufficient condition for separability.

However, let us make an amended choice $\beta'$: take the observables
$\beta$ and apply a rotation $U$ for the observables of particle 1 around
the $y$-axis over 45 degrees, i.e.  take $(A, A', A'')_{\beta'} = (
U\sigma_zU^\dagger , \sigma_y , U \sigma_x U^\dagger)$ and
$(B,B',B'')_{\beta'} = (\sigma_z, \sigma_y, \sigma_x)$
; and $\gamma'$: take the observables of choice $\gamma$ and
apply rotation $U$  on the observables for particle 1   (i.e., over 45 degrees around the
$y$-axis) followed up by rotation $V$ over 45 degrees around the $z$-axis on the same observables,   in other
words:  $(A,A',A''))_{\gamma'} = (VU\sigma_zU^\dagger V^\dagger,VU
\sigma_xU^\dagger V^\dagger ,VU \sigma_yU^\dagger V^\dagger)$ and
$(B,B',B'')_{\gamma'} = (\sigma_z, \sigma_x, \sigma_y)$.

  The choice $\alpha$, $\beta'$ and $\gamma'$  gives for the above arbitrary  pure
state $\ket{\Psi}$:
\begin{align} |ad|   &= |bc| , \\
|   (a+c)  (b-d) |   &=   |(a-c)  (b+d) |,\\
 | (a+ i c) (b-id) | &= | (a -ic)(b+id)|. \end{align}
A tedious but straightforward calculation shows that these
equations are fulfilled \emph{only} if $ad =bc$, i.e., if
$\ket{\Psi}$ is separable.  Hence, by  measuring the observables
in the directions indicated by  the choice $\alpha$, $\beta'$ and $\gamma'$,
the inequalities  (\ref{mixsep2}) do
 provide a necessary and sufficient criterion for separability for
 pure states.  We have  not been able to check whether this result extends
 to mixed states.

\section{Conclusion}
It has been shown that for two spin-${1}/{2}$ particles (qubits)
and orthogonal spin components  quadratic separability inequalities hold
that impose much tighter bounds on the correlations in separable
states than the traditional Bell-CHSH inequality.
\forget{ This result provides
a quadratic entanglement witness that improves on all
existing witnesses derived from Bell inequalities.}
In fact, the quadratic inequalities (\ref{mixsep2}) are so strong that
their validity for all orthogonal bases is a necessary and sufficient condition
for separability of all states, pure or mixed, and a subset of these
inequalities for just three orthogonal bases (giving six inequalities) is a necessary and sufficient
condition for the separability of all pure states.
Furthermore, the orientation of the measurement basis is shown to be irrelevant,
which ensures that no shared reference frames needs to be established
between the measurement apparata for each qubit.

The quadratic inequalities
 (\ref{mixsep2}) have been shown to be stronger
 than both the fidelity criterion and the linear and non-linear entanglement witnesses based on LOOs as given in \cite{yu, nonlinear}.
Experimental tests for entangled states using orthogonal directions can
therefore be considerably strengthened by means of the quadratic
inequalities (\ref{mixsep2}). As we will discuss
elsewhere in more detail \cite{future}, these inequalities provide tests of
entanglement that are much more robust against noise than many
alternative criteria. There we will also extend the analysis
to the $N$-qubit case by generalizing the method of
section 4 to more than two qubits.

Furthermore, we have argued that these quadratic Bell inequalities do
not hold in LHV theories.  This provides a more general example of
the fact first discovered by Werner, i.e.,  that some entangled
states do allow an LHV reconstruction for all correlations in a
standard Bell experiment.
What is more, there appears to be a  `gap' between the
correlations that can be obtained by separable quantum states and
those obtainable by LHV models. This non-equivalence between the
correlations obtainable from separable quantum states and from LHV
theories means that, apart from the question raised and answered
by Bell (can the predictions of quantum mechanics be reproduced by
an LHV theory?)
  it is also interesting to ask whether separable quantum states
can reproduce the predictions of an LHV theory. The answer, as we
have seen, is negative:  quantum theory generally needs entangled
states even in order to reproduce the classical correlations of
such an LHV theory. In fact, as we will show in forthcoming work
\cite{future}, the gap between the correlations allowed by local
hidden-variable theories and those achievable by separable quantum
states increases exponentially with the number of particles.


\section*{Appendices}

\noindent
{\bf  Appendix A} ---
Here we prove that any pure two-qubit state
satisfying (\ref{newa}) must be separable.  By the bi-orthogonal
decomposition theorem, and following Ref. \cite{GISIN91}, any pure
state can be written in the form $ \ket{\Psi} =$ \mbox{$r\ket{\up\down} -
s\ket{\down \up}$}, with $r,s \geq 0$, $r^2 +s^2 =1$. For this
state $\av{\vec{a}\cdot \vec{\sigma} \otimes \vec{b}\cdot
\vec{\sigma}}_{\Psi} =-a_z b_z - 2 rs \left(a_x b_x +
a_yb_y\right)$, etc.
 Using this and choosing $\vec{a} =(0, \,0,\, 1),~\vec{a'}=(1, \,0,\, 0)$ and $
\vec{b} = (\sin\beta,\,0,\,\cos\beta),~\vec{b'} = (-\cos
\beta,\,0 ,\,\sin \beta)
$ we obtain $\av{AB' + A'B}^2 + \av{AB - A' B'}^2
= (1 + 2rs)^2$. If (\ref{new}) holds, this
expression is smaller than or equal to 1, and
 it follows that $rs=0$, i.e., the state $\ket{\Psi}$  is not
entangled.

\noindent
{\bf Appendix B} --- Here we provide further examples of entangled states that satisfy
the Bell-CHSH inequalities (\ref{1}) for all observables in the standard
Bell experiment. First note \cite{H3} that any two-qubit state can
be written in the form $\rho = \frac{1}{4} ( \1 \otimes \1 +
\vec{r}\cdot \vec{\sigma} \otimes \1 + \1\otimes \vec{s}\cdot
\vec{\sigma} + \sum_{ij=1}^3 t_{ij}\, \sigma_i \otimes \sigma_j )
$, where $ \vec{r} = \Tr\rho (\vec{\sigma}\otimes \1)$, $\vec{s} =
\Tr \rho (\1\otimes \vec{\sigma}) $  and $t_{ij} = \Tr \rho
(\sigma_i\otimes \sigma_j)$. By employing the freedom of choosing
local coordinate frames at both sites separately, we can bring the
matrix $(t_{ij})$ to diagonal form \cite{H2}, i.e., $t=
\mbox{diag~}(t_{11}, t_{22}, t_{33})$, and arrange that $t_{ii}
\geq 0$. Furthermore, since the labelling of the coordinate axes
is arbitrary, we can also pick an ordering such that $t_{11} \geq
t_{22}\geq t_{33}$.

Now let $\alpha, \alpha', \beta, \beta'$ denote two pairs of
arbitrary spin observables, for particle 1 and 2 respectively,
$\alpha = \vec{\alpha}\cdot \vec{\sigma} \otimes \1$, $ \beta =
\1 \otimes \vec{\beta} \cdot \vec{\sigma}$ and similar for the
primed observables. It is easy to see that
 the maximum of
 $|\av{\cal \alpha\beta + \alpha\beta' + \alpha'\beta - \alpha'\beta'}_\rho|$
 for all choices of observables  will
 be attained  by taking the vectors $\vec{\alpha}, \vec{\alpha'},
\vec{\beta},\vec{\beta'}$ coplanar \footnote{Here `coplanar' refers
to a single plane in the local frames of reference. Since these
frames may have a different orientation, this does not
necessarily refer to a single plane in real space.}, and in
fact, in the plane spanned by the two eigenvectors of $t$ with the
largest eigenvalues, i.e., $t_{11}$ and $t_{22}$. As shown by
Ref.~\cite{H3}, this maximum is $
\max_{\alpha,\beta,\alpha',\beta'}|\av{ \alpha\beta +
\alpha\beta' + \alpha'\beta - \alpha'\beta'}_\rho|= 2
\sqrt{t_{11}^2 + t_{22}^2 }$. Thus $\rho$ will satisfy all
Bell-CHSH inequalities if $t^2_{11} + t^2_{22} \leq 1$, which is
the necessary and sufficient condition for the existence of an LHV
model \cite{zukowskibrukner}.

Now  consider  the maximum of $\av{ AB - A'B'}_{\rho}^2  + \av{AB' +
A'B}_\rho^2$, with $A\perp A'$ and $B \perp B'$. Clearly, these
spin observables  should be chosen in the same plane as before,
spanned by the eigenvectors corresponding to $t_{11}$ and
$t_{22}$. As mentioned in the text, the expression is invariant
under rotations of $A,A'$ or $B,B'$ in this plane. Choosing
$A = B =\sigma_x $, $A'= - B' = \sigma_y$ the maximum is equal to
$\max_{A\perp A', B\perp B'} \av{ AB - A'B'}^2 + \av{AB' + A'B}^2
 = (t_{11} + t_{22})^2. $
Clearly,  state $\rho$ will be  both entangled and satisfy
all Bell-CHSH inequalities for all observables
(and thus allow for an LHV description)  if $   t_{11} +
t_{22}>1~~~\mbox{and}~~~ t^2_{11} + t^2_{22} \leq 1$.

\noindent
{\bf Appendix C} ---
Here we will prove that any state $\rho$ that satisfies the inequalities (\ref{mixsep2}) for
all orthogonal triples $A, A', A''$, and  $B, B', B''$ must be separable (the converse has already been proven above).

We proceed from the well-known Peres-Horodecki lemma  \cite{PPT}
that a state of two qubits is separable iff $\rho^{\rm PT}\geq 0 \label{PT}$ where 'PT' denotes partial transposition.
Equivalently, the state is entangled iff, for all pure states
$\ket{\Psi}$: \beq \bra{\Psi} \rho^{\rm PT} \ket{\Psi} =\Tr
\rho^{\rm PT} \ket{\Psi}\bra{\Psi} = \Tr \rho
(\ket{\Psi}\bra{\Psi})^{PT} \geq 0 \label{PT'}.\enq
We shall show that (\ref{PT'}) holds whenever $\rho$ obeys (19). Indeed, according to the biorthonormal decomposition theorem (cf. \cite{GISIN91}), we can find bases
$\ket{\up}, \ket{\down}$ on ${\cal H}_1$ and $\ket{\up},\ket{\down}$ on ${\cal
H}_2$ such that \mbox{$\ket{\Psi} =   \sqrt{p} \ket{\up\down}  + \sqrt{1-p} \ket{\down\up}$}.
Choosing these bases to be the eigenvectors of $A''$ and $B''$ respectively, we
thus find \begin{align}
 \ket{\Psi}\bra{\Psi} &= \half \tilde{I} + (p-\half)\tilde{ Z} +
 \sqrt{p(1-p)} \tilde{X}, \nn\\
\ket{\Psi}\bra{\Psi}^{PT} &= \half \tilde{I} + (p-\half)\tilde{
Z} +
 \sqrt{p(1-p)} {X} .\end{align}
Hence \beq \bra{\Psi}\rho^{PT} \ket{\Psi} = \half \av{\tilde{I}} +
(p-\half) \av{\tilde{Z}} + \sqrt{p(1-p)} \av{X} ,\enq where the
last two terms can be bounded by a  Schwartz inequality to yield
\beq  | (p-\half) \av{\tilde{Z}} + \sqrt{p(1-p)} \av{X}| \leq
\half \sqrt{\av{\tilde{Z}}^2  + \av{X}^2} \enq and we find $ \bra{\Psi}\rho^{PT} \ket{\Psi} \geq  \half \av{\tilde{I}}  -
\half \sqrt{\av{\tilde{Z}}^2  + \av{X}^2}$.  But (19) demands $ \av{X}^2_\rho + \av{\tilde{Z}}_\rho^2  \leq
 \av{\tilde{I}}^2_\rho $ from which it follows that \beq \bra{\Psi}\rho^{PT} \ket{\Psi} \geq 0\enq so
that the state $\rho$ is separable.


\begin{thebibliography}{99}
\bibitem{entanglement} R. Horodecki, {\it et al.}. ArXiv: quant-ph/0702225 (2007).
\bibitem{nielsen}M. A. Nielsen,  I. L. Chuang, Quantum Computation and Quantum Information, Cambridge University Press, Cambridge, 2000.
\bibitem{BELL64} J.S. Bell, Physics 1 (1964) 195.
\bibitem{GISIN91} N.~Gisin, Phys. Lett. A 154 (1991) 201.
\bibitem{H3} R. Horodecki, P. Horodecki, M. Horodecki, Phys. Lett. A 200 (1995) 340.
\bibitem{volz} J. Volz, {\emph et al.}, Phys. Rev. Lett.  96 (2006) 030404. R.M. Stevenson,  {\emph et al.}, Nature  439 (2006) 179.
\bibitem{hefei}  S. Yu, J.-W. Pan,  Z.-B. Chen, Y.-D. Zhang, Phys. Rev. Lett.  91 (2003) 217903.
\bibitem{sackett} C.A.~Sackett, \emph{et al.}, \forget{ D.~Kielpinski, B.E.~King, C.~Langer, V.~Meyer,
C.J.~Myatt, M.~Rowe, Q.A.~Turchette, W.M.~Itano, D.J.~Wineland,
C.~Monroe,} Nature 404 (2000) 256. M.~Seevinck, J.~Uffink, Phys.  Rev.  A  65 (2001) 012107.
\bibitem{yu}S. Yu, N.-L. Liu, Phys. Rev. Lett. 95 (2005) 150504.
\bibitem{nonlinear} O. G\"uhne, M. Mechler, G. T\'oth, P. Adam, Phys. Rev. A 74 (2006) 010301.
\bibitem{zhang}C-J. Zhang \emph{et al.}, ArXiv: quant-ph/0705.1832 (2007).
\bibitem{WERNER89} R.F. Werner, Phys. Rev. A 40 (1989) 4277.
\bibitem{chsh} J.F. Clauser, M.A. Horne, A. Shimony, R.A. Holt, Phys. Rev. Lett. 26 (1969) 880.
\bibitem{CIRELSON} B.S.~Cirel'son, Lett. Math. Phys. 4 (1980) 93.
\bibitem{LANDAU}L.J.~Landau,  Phys. Lett. A 120 (1987) 54.
\bibitem{UFFINK02}  J.~Uffink, Phys.  Rev.  Lett. 88 (2002) 230406.
\bibitem{TOTHSEEV}G. T\'oth, \emph{et al.}, Phys. Rev. A 72 (2005) 014101.
\bibitem{popescu}S. Popescu and D. Rohrlich, Phys. Lett. A 169 (1992) 411.
\bibitem{wernerwolf} R.F. Werner, M.M. Wolf, Phys. Rev. A, 61 (2000)  062102.
\bibitem{seevuff2007} M. Seevinck, J. Uffink, Phys. Rev. A 76 (2007) 042105.
\bibitem{POPESCUROHRLICH} S. Popescu, D. Rohrlich, Phys. Lett. A 166 (1992) 293.
\bibitem{roy} S.M. Roy, Phys. Rev. Lett.  94 (2005) 010402.
\bibitem{BELL} J.S. Bell, Rev. Mod. Phys. 38 (1966) 447;  J.S. Bell, Foundations of quantum mechanics,
 Proceedings of the International School of  Physics `Enrico Fermi', New York, Academic, 1971, pp. 171-181.
\bibitem{acin} A. Acin, N. Gisin, B. Toner, Phys. Rev. A 73 (2006) 062105.
\bibitem{wernerwolf2} R.F. Werner, M.M. Wolf, Phys. Rev. A 64 (2001) 032112.
\bibitem{zukowskibrukner} M. \.Zukowski,  \v C. Brukner, Phys. Rev. Lett. 88 (2002) 210401.
\bibitem{fine}A. Fine, Phys. Rev. Lett. 48 (1982) 291.
\bibitem{pitowsky} I. Pitowsky, Quantum Probability--Quantum Logic, Springer, Berlin, 1989.
\bibitem{footnote}Note that experiments with more general measurement scenarios
(e.g., collective, sequential
or postselected measurements) might still produce
correlations incompatible with any LHV model.
 However, we will not discuss this issue.
\bibitem{PPT} A. Peres,  Phys. Rev. Lett. 77 (1996) 1413 (1996);  M. Horodecki, P. Horodecki,  R. Horodecki, Phys. Lett. A 223 (1996) 1.
\bibitem{future} M. Seevinck and J. Uffink, in preparation.
\bibitem{H2}R. Horodecki, M. Horodecki, Phys. Rev. A 54 (1996) 1838.
%


\end{thebibliography}
\end{document}